\documentclass[twocolumn]{IEEEtran} 
   \usepackage{epsfig}
   \usepackage{color}
   \usepackage{latexsym}
   \usepackage{pslatex}

\newtheorem{lemma}{Lemma}
\newtheorem{theorem}{Theorem}

\begin{document}

\title{Finding Matching Initial States for Equivalent NLFSRs in the Fibonacci and the Galois
Configurations}
\author{Elena Dubrova \\
Royal Institute of Technology (KTH), Electrum 229, 164 46 Kista, Sweden\\
dubrova@kth.se}

\maketitle

\begin{abstract}
In this paper, a mapping between initial states of the Fibonacci and the Galois
configurations of NLFSRs is established. 
We show how to choose initial states 
for two configurations so that the resulting output sequences are equivalent.
\end{abstract}


\section{Introduction} 

Non-Linear Feedback Shift Registers (NLFSR) are a generalization of 
Linear Feedback Shift Registers (LFSRs) in which a current state is 
a non-linear function of the previous state~\cite{Golomb_book}. While the theory behind LFSRs is
well-understood, many fundamental questions related to NLFSRs remain open.

The interest in NLFSRs is motivated by their ability to generate
pseudo-random sequences which are hard to break with existing cryptanalytic
methods~\cite{Ca05}. 
A common approach for encrypting confidential information
is to use a {\em stream cipher} which combines plain text bits with a pseudo-random bit sequence~\cite{robshaw94stream}.
The resulting encrypted information can be transformed back into its original form only by an authorized 
user possessing the cryptographic key.
While LFSRs are widely used in testing and simulation~\cite{AbBF94}, for cryptographic applications their pseudo-random sequences are not secure. The structure of an $n$-bit LFSR can be easily deduced by observing $2n$ consecutive bit of its sequence~\cite{Ma69}.
Contrary, an adversary might need $2^n$ bits of a sequence to determine the structure of the $n$-bit NLFSR which generates it~\cite{DuTT08}.
A number of NLFSR-based stream
ciphers for RFID and smartcards applications have been proposed,
including Achterbahn~\cite{GaGK07},
Grain~\cite{hell-grain}, Dragon~\cite{CHM05}, Trivium~\cite{canniere-trivium},
VEST~\cite{cryptoeprint:2005:415},
and the cipher~\cite{GaGK06}.

Similarly to LFSRs, an NLFSR can be implemented either in the Fibonacci or in the Galois hardware configuration.
In the former, the feedback is applied to the last
bit of the register only, while in the latter the
feedback can potentially be applied to every bit. 
The depth of circuits implementing feedback
functions in a Galois configuration is usually smaller than the one in the equivalent Fibonacci configuration~\cite{Du08}.
This makes the Galois configuration more attractive
for stream ciphers where high throughput is important.
For example, by re-implementing the NLFSR-based stream cipher Grain~\cite{hell-grain} from
the original Fibonacci to the Galois configuration, one can double the throughput with no penalty in area or power~\cite{Sh09}.

In~\cite{Du08} it has been shown how to transform a Fibonacci NLFSR into an equivalent Galois NLFSR.
While the resulting NLFSRs
generate the same sets of output sequences, they follow different sequences of states
and normally start from a different initial state. The relations between sequences of states and
between 
initial states of two configurations are studied in this paper.
One reason for studying the relation between sequences of states is that some NLFSR-based 
stream ciphers use not only the output of
an NLFSR, but also several other bits of its state 
to produce a pseudo-random sequence. If a Fibonacci to Galois transformation is applied to
an NLFSR-based stream cipher,
it is important to know which
bits of the state are affected by the transformation
in order to preserve the original algorithm. 
Changing the algorithm is likely to influence the security of a cipher.
For the same reason, 
we need to map the 
secret key and the initial value (IV) of the original cipher into the corresponding ones of
the transformed cipher. Finally,  knowing which initial state of the
Galois configuration matches a given initial state of the Fibonacci configuration 
makes possible validating the equivalence of two configurations by simulation.


The paper is organized as follows. Section~\ref{back} gives an introduction to 
NLFSRs and describes the Fibonacci to Galois transformation.
In Section~\ref{main1},
we study a relation between the sequences of states 
generated by two equivalent NLFSRs. Section~\ref{main2} shows how to compute the initial state for the Galois 
configuration 
which matches a given initial state of the Fibonacci configuration.
Section~\ref{conc} concludes the paper and
discusses open problems.

\section{Background} \label{back}

In this section, we give an introduction to NLFSRs and briefly describe
the transformation from the Fibonacci to the Galois configuration. 
For more details, the reader is referred to~\cite{Du08}.

\subsection{Definition of NLFSRs}

A {\em Non-Linear Feedback Shift Register (NLFSR)} consist of $n$
binary storage elements, called {\em bits}. 
Each bit $i \in \{0,1,\ldots,n-1\}$ has an associated {\em state variable} $x_i$ 
which represents the current value of the bit $i$ and a {\em feedback function} 
$f_i: \{0,1\}^n \rightarrow \{0,1\}$ which determines how the value of $i$ is updated. 
For any $i \in \{0,1,\ldots,n-1\}$, $f_i$ depends on $x_{(i+1) mod~n}$
and a subset of variables from the set  $\{x_0, x_1, \ldots, x_i\}$.

A {\em state} of an NLFSR is an ordered set of values of its
state variables $(x_0, x_1, \ldots, x_{n-1})$.
At every clock cycle, 
the next state is determined from the current state 
by updating the values of all bits simultaneously
to the values of the corresponding $f_i$'s.
The {\em output} of an NLFSR is the value of its 0th bit.
The {\em period} of an NLFSR is the length of the longest
cyclic output sequence it produces. 

If for all  $i \in \{0,1,\ldots,n-2\}$ the feedback functions are of type $f_i = x_{i+1}$, we call
an NLFSR the {\em Fibonacci} type. Otherwise, we call an NLFSR the {\em Galois} type.

Two NLFSRs are {\em equivalent} if their sets of output sequences are equivalent.

Feedback functions of NLFSRs are usually represented using the algebraic normal form.
The {\em algebraic normal form (ANF)} of a Boolean function $f:
\{0,1\}^n \rightarrow \{0,1\}$ is a polynomial in $GF(2)$ of type
\[
f(x_0, \ldots, x_{n-1}) = \sum_{i=0}^{2^n-1}  c_i \cdot 
x_0^{i_0} \cdot x_1^{i_1} \cdot \ldots \cdot x_{n-1}^{i_{n-1}},
\]
where $c_i \in \{0,1\}$ and $(i_0 i_1 \ldots i_{n-1})$ is the binary
expansion of $i$ with $i_0$ being the least significant bit. 
Throughout the paper, we call a term of the ANF a {\em product-term}.

\subsection{The transformation from the Fibonacci to the Galois configuration} \label{prel}

Let $f_i$ and $f_j$ be feedback functions of bits $i$ and $j$ of an
$n$-bit NLFSR, respectively. The operation {\em shifting}, denoted by $f_i
\stackrel{P}{\rightarrow} f_j$, moves a set of product-terms $P$
from the ANF of $f_i$ to the ANF of $f_j$.
The index of each variable $x_k$ of each product-term in $P$
is changed to $x_{(k-i+j)~\mbox{\small{mod}}~n}$.  

The {\em terminal bit} $\tau$ of an $n$-bit NLFSR is the bit with the maximal index
 which satisfies the following condition: 
\[
\mbox{For all bits $i$ such that $i < \tau$, $f_i$ is of type $f_i = x_{i+1}$.}
\]

An $n$-bit NLFSR is {\em uniform} if the following two condition hold: 
\begin{enumerate}
\item[(a)] all its feedback functions are {\em singular} functions of type
\[
f_i(x_0,\ldots,x_{n-1}) = x_{(i+1) mod~n} \oplus g_i(x_0,\ldots,x_{n-1}),
\]
where $g_i$ does not depend on $x_{(i+1) mod~n}$,
\item[(b)] for all its bits $i$ such that $i > \tau$, 
the index of every variable of $g_i$ is not larger than $\tau$.
\end{enumerate}

\begin{theorem} \cite{Du08}
Given a uniform NLFSR with the terminal bit $\tau$, a shifting $g_{\tau} \stackrel{P}{\rightarrow} g_{\tau'}$,
 $\tau' < \tau$, results in an equivalent NLFSR if the transformed NLFSR is uniform as well.
\end{theorem}

\section{The Relation Between Sequences of States}  \label{main1}

Although a Fibonacci NLFSR and a Galois NLFSR can generate the same output sequence,
they follow different sequences of states.  
Therefore, in order to generate the same output sequence, they normally have to be set to different initial states. 
In this section we study the relation between sequences of states 
produced by two equivalent NLFSRs and derive a basic property which will be used to prove of the
main result of the paper.

Let $s = (s_0,s_1,\ldots,s_{n-1})$ be a state of an NLFSR, $s_i \in \{0,1\}$.
Throughout the paper, we use $g_i(s)$ to denote the value of the function $g_i$ evaluated for the vector $s$.
We also use $g_i|_{+m}$  to denote the function obtained from the function $g_i$ by increasing
indexes of all variables of $g_i$ by $m$. 
For example, if $g_1 = x_1 \cdot x_2 \oplus x_3$, then 
$g_1|_{+2} = x_3 \cdot x_4 \oplus x_5$. 
To simplify the exposition, we do not list variables of a function explicitly
if it does not cause any ambiguity, i.e. in the previous example we
wrote $g_1$ instead of $g_1(x_1,x_2,x_3)$.


\begin{lemma} \label{l_main}
Let $N_1$ be an $n$-bit uniform NLFSR with the terminal bit $\tau$, $0 < \tau \leq n-1$, 
which has the feedback function of type
\[
f_{\tau} = x_{(\tau+1) mod~n} \oplus g_{\tau} \oplus p_{\tau}
\]
and let $N_2$ be an equivalent uniform NLFSR obtained from $N_1$
by shifting from $\tau$  to $\tau-1$ the set of product-terms represented by the function $p_{\tau}$.

If $N_1$ is initialized to a state $s  = (s_0,s_1,\ldots,s_{n-1})$  
and 
$N_2$ is initialized to the state $(s_0,s_1,\ldots,s_{\tau-1},r_{\tau},s_{\tau+1},\ldots,s_{n-1})$,
where
\begin{equation} \label{er1}
r_{\tau} = s_{\tau}  \oplus p_{\tau}|_{-1}(s) 
 \end{equation} 
then they generate sequences of states which differ in the bit $\tau$ only.
\end{lemma}
{\bf Proof:} 
Suppose that $N_1$ is initialized to a state 
$s = (s_0,s_1,\ldots,s_{n-1})$ and $N_1$ is initialized to a state $r = (r_0,r_1,\ldots,$   $r_{n-1})$,
such that  $r_i = s_i$ for all $i$ except  $i = \tau$ and $r_{\tau}$ is given by (\ref{er1}).

On one hand, for $N_1$, the next state is $s^+ = (s^+_0,s^+_1,\ldots,s^+_{n-1})$ such that
\[
\begin{array}{l}
s^+_{n-1} = s_0 \oplus g_{n-1}(s_1,s_2,\ldots,s_{\tau-1}) \\
\ldots \\
s^+_{\tau} = s_{\tau+1} \oplus g_{\tau}(s_0,s_1,\ldots,s_{\tau-1})  \oplus p_{\tau}(s_1,s_2,\ldots,s_{\tau})   \\
s^+_{\tau-1} = s_{\tau} \\
\ldots \\
s^+_0 = s_1. \\
\end{array}
\]
Note that, since $N_1$ is uniform, the functions $g_{n-1}, g_{n-2}, \ldots, g_{\tau}$ may only depend on variables 
with indexes between 0 to $\tau$. Furthermore, $g_{n-1}, g_{n-2}, \ldots, g_{\tau}$ cannot depend on the variable $x_\tau$, 
since otherwise $N_2$ would not be uniform after shifting. For the same reason, the function $p_{\tau}$ cannot depend on 
the variable $x_0$.

On the other hand, for $N_2$, the next state is $r^+ = (r^+_0,r^+_1,\ldots,r^+_{n-1})$, where
\[
\begin{array}{l}
r^+_{n-1} = r_0 \oplus g_{n-1}(r_1,r_2,\ldots,r_{\tau-1}) \\
\ldots \\
r^+_{\tau} = r_{\tau+1} \oplus g_{\tau}(r_0,r_1,\ldots,r_{\tau-1})  \\
r^+_{\tau-1} = r_{\tau} \oplus p_{\tau}|_{-1}(r_0,r_1,\ldots,r_{\tau-1}) \\
r^+_{\tau-2} = r_{\tau-1} \\
\ldots \\
r^+_0 = r_1. \\
\end{array}
\]
By substituting $r_i = s_i$ for all $i$ except  $i = \tau$, we get:
\[
\begin{array}{l}
r^+_{n-1} = s_0 \oplus g_{n-1}(s_1,s_2,\ldots,s_{\tau-1}) \\
\ldots \\
r^+_{\tau} = s_{\tau+1} \oplus g_{\tau}(s_0,s_1,\ldots,s_{\tau-1})  \\
r^+_{\tau-1} = r_{\tau} \oplus p_{\tau}|_{-1}(s_0,s_1,\ldots,s_{\tau-1}) \\
\ldots \\
r^+_0 = s_1. \\
\end{array}
\]
By substituting $r_{\tau}$ by (\ref{er1}), we get
\[
\begin{array}{ll}
r^+_{\tau-1} & = s_{\tau}  \oplus p_{\tau}|_{-1}(s_0,s_1,\ldots,s_{\tau-1}) \oplus p_{\tau}|_{-1}(s_0,s_1,\ldots,s_{\tau-1}) \\
 & = s_{\tau}. \\
\end{array}
\]
So, the next state of $N_2$ is
\[
\begin{array}{l}
r^+_{n-1} = s^+_{n-1} \\
\ldots \\
r^+_{\tau} = s_{\tau+1} \oplus g_{\tau}(s_0,s_1,\ldots,s_{\tau-1})  \\
r^+_{\tau-1} = s^+_{\tau-1} \\
\ldots \\
r^+_0 = s^+_1 \\
\end{array}
\]
i.e. the next states of $N_1$ and $N_2$ can potentially differ only the bit position $\tau$.

In order to extend this conclusion to a sequence of states, it
remains to show that the resulting $r^+_{\tau}$ can be expressed according to (\ref{er1}).
From 
\[
s^+_{\tau} = s_{\tau+1} \oplus g_{\tau}(s_0,s_1,\ldots,s_{\tau-1})  \oplus p_{\tau}(s_1,s_2,\ldots,s_{\tau})   \\
\]
we can derive
\[
s_{\tau+1} = s^+_{\tau} \oplus g_{\tau}(s_0,s_1,\ldots,s_{\tau-1})  \oplus p_{\tau}(s_1,s_2,\ldots,s_{\tau}).  
\]
Substituting it to the expression of $r^+_{\tau}$ above and eliminating the double occurrence of 
$g_{\tau}(s_0,s_1,\ldots,s_{\tau-1})$, we get
\[
r^+_{\tau} = s^+_{\tau} \oplus p_{\tau}(s_1,s_2,\ldots,s_{\tau})
\]
Since $p_{\tau}(s_1,s_2,\ldots,s_{\tau}) = p_{\tau}|_{-1}(s^+_0,s^+_1,\ldots,s^+_{\tau-1})$, we get
\[
r^+_{\tau} = s^+_{\tau}  \oplus p_{\tau}|_{-1}(s^+) 
\]
\begin{flushright}
$\Box$
\end{flushright}

As an example, consider the following 4-bit NLFSR $N_1$:
\[
\begin{array}{l}
f_3 = x_0 \oplus x_1 \\
f_2 = x_3 \oplus x_1 \oplus x_0 x_1 \\
f_1 = x_2 \\
f_0 = x_1. 
\end{array}
\]
which has the period 15. Suppose we shift the product term $x_1$ from the bit 2 to the bit 1. Then
we get the following equivalent NLFSR  $N_2$:
\[
\begin{array}{l}
f_3 = x_0 \oplus x_1 \\
f_2 = x_3 \oplus x_0 x_1 \\
f_1 = x_2 \oplus x_0 \\
f_0 = x_1. 
\end{array}
\]
The sequences of states of $N_1$ and $N_2$ are shown in the 1st and 2nd columns of Table~\ref{ex}.
The initial states of $N_1$ and $N_2$ are  $(s_3 s_2 s_1 s_0) = (0001)$ and $(r_3 r_2 r_1 r_0) = (0101)$, respectively.
According to Lemma~\ref{l_main}, we have $r_0 = s_0$, $r_1 = s_1$, $r_2 = s_2 \oplus s_0$, and $r_3 = s_3$. 
As we can see, these sequences differ in the bit 2 only, which is the terminal
bit of $N_1$.

\begin{table}[t]
\begin{center} 
\caption{Sequences of states of three equivalent 4-bit NLFSRs.} \label{ex}
\begin{tabular}{|c|c|c|} \hline
\multicolumn{2}{|c|}{Galois} & Fibonacci \\ \hline
NLFSR $N_1$ & NLFSR $N_2$ & NLFSR $N_3$ \\ 
$x_3 x_2 x_1 x_0$  & $x_3 x_2 x_1 x_0$ & $x_3 x_2 x_1 x_0$ \\ \hline 
0 0 0 1  		&			0 1 0 1  &   0 0 0 1 \\
1 0 0 0 		&			1 0 0 0  &   1 0 0 0 \\
0 1 0 0  		&			0 1 0 0  &   0 1 0 0 \\
0 0 1 0  		&			0 0 1 0  &   1 0 1 0 \\
1 1 0 1  		&			1 0 0 1  &   1 1 0 1 \\
1 1 1 0  		&			1 1 1 0  &   0 1 1 0 \\
1 0 1 1  		&			1 1 1 1  &   1 0 1 1 \\
0 1 0 1  		&			0 0 0 1  &   0 1 0 1 \\
1 0 1 0  		&			1 0 1 0  &   0 0 1 0 \\
1 0 0 1  		&			1 1 0 1  &   1 0 0 1 \\
1 1 0 0  		&			1 1 0 0  &   1 1 0 0 \\
0 1 1 0  		&			0 1 1 0  &   1 1 1 0 \\
1 1 1 1  		&			1 0 1 1  &   1 1 1 1 \\
0 1 1 1  		&			0 0 1 1  &   0 1 1 1\\
0 0 1 1  		&			0 1 1 1  &   0 0 1 1 \\ \hline 
\end{tabular}
\end{center}
\end{table}

The following property follows trivially from Lemma~\ref{l_main}.

\begin{lemma} \label{l_e}
Let $N_1$ be an $n$-bit uniform NLFSR with the terminal bit $\tau$, $0 < \tau \leq n-1$, 
which has the feedback function of type
\[
f_{\tau} = x_{(\tau+1) mod~n} \oplus g_{\tau} \oplus p_{\tau}
\]
and let $N_2$ be an equivalent uniform NLFSR obtained from $N_1$
by shifting from $\tau$ to $\tau-1$ the set of product-terms represented by the function $p_{\tau}$.

If $N_1$ is initialized to a state $s  = (s_0,s_1,\ldots,s_{n-1})$  
and 
$N_2$ is initialized to the state $(s_0,s_1,\ldots,s_{\tau-1},r_{\tau},s_{\tau+1},\ldots,s_{n-1})$,
such that
\begin{equation} \label{er}
r_{\tau} = s_{\tau}  \oplus p_{\tau}|_{-1}(s),
 \end{equation} 
then $N_1$ 
and $N_2$ generate the same output sequence.
\end{lemma}

As an example, consider the sequences of states of NLFSRs
$N_1$ and $N_2$ 
shown in the 1st and 2nd columns of Table~\ref{ex}.
Since their initial states $(0001)$ and $(0101)$ agree with Lemma~\ref{l_e},
 $N_1$ and $N_2$ generate the same output sequence $100010110100111$.

\section{The Mapping Between Initial States} \label{main2}

This section presents the main result of the paper.

\begin{theorem} \label{init_state}
Let $N_F$ be an $n$-bit Fibonacci NLFSR 
and $N_G$ be an equivalent uniform Galois NLFSR with the terminal bit $0 \leq \tau < n-1$
and the feedback functions of type
\begin{equation} \label{eg}
\begin{array}{l}
f_{n-1} = x_0 \oplus g_{n-1} \\
f_{n-2} = x_{n-1} \oplus g_{n-2} \\
\ldots \\
f_{\tau}= x_{\tau+1} \oplus g_{\tau} \\
f_{\tau-1} = x_{\tau} \\
\ldots\\
f_0 = x_1.
\end{array}
\end{equation}

If $N_F$ is initialized to a state 
$s = (s_0,s_1,\ldots,s_{n-1})$
and $N_G$ is 
initialized to the state $(s_0,s_1,\ldots,s_{\tau},r_{\tau+1},r_{\tau+2},\ldots,r_{n-1})$
such that 
\[
r_i  = s_i \oplus g_{i-1}(s) \oplus g_{i-2}|_{+1}(s) \oplus \ldots \oplus g_{\tau}|_{+i-\tau-1}(s) 
\]
for all $i ‌ \in \{n-1,n-2,\ldots,\tau+1\}$,
then $N_F$ and $N_G$ generate the same output sequence.
\end{theorem}
{\bf Proof:} 
From the definition of shifting, we can conclude that if, after the transformation, the Galois NLFSR 
has feedback functions of type~(\ref{eg}),
then, the feedback function of the $n-1$th bit of the original Fibonacci NLFSR is of type:
\[
f'_{n-1} = x_0 \oplus  g_{n-1} \oplus g_{n-2}|_{+1} \oplus g_{n-3}|_{+2} \oplus \ldots \oplus g_{\tau}|_{+n-1-\tau}.
\]

Any uniform Galois NLFSR can be obtained by 
first shifting all product-terms of the original Fibonacci NLFSR
but the ones represented by $g_{n-1}$ from the bit $n-1$ to the bit $n-2$, then
shifting all product-terms 
but the ones represented by $g_{n-2}$ from the bit $n-2$ to the bit $n-3$, etc.,
i.e. using a sequence of $n-1-\tau$ shiftings by one bit.
This means that, at each step, the set of product-terms represented by the function 
\begin{equation} \label{pr}
p_{n-1-i} = g_{n-1-i-1}|_{+1} \oplus g_{n-1-i-2}|_{+2} \oplus \ldots \oplus g_{\tau}|_{+n-1-i-\tau}
\end{equation} 
is shifted from the bit $n-1-i$ to the bit  $n-1-i-1$, for $i \in \{0,1,\ldots,n-1-\tau-1 \}$.
Furthermore, for each $i \in \{0,1,\ldots,n-1-\tau-1 \}$, 
by Lemma~\ref{l_e}, if the NLFSR before shifting is initialized to some state $s'$  
and the NLFSR after shifting is initialized to the state where the bit $n-1-i$ has
the value $s_{n-1-i} \oplus  p_{n-1-i}|_{-1}(s')$ and all other bits have the same values 
as the corresponding bits of $s'$, 
then two NLFSRs generate the same output sequence.

Therefore, we can conclude that if the original Fibonacci NLFSR $N_F$ is initialized to the state 
$s = (s_0,s_1,\ldots,s_{n-1})$
and the NLFSR $N_G$ obtained using the sequence of $n-1-\tau$ shiftings by one bit described above is 
initialized to the state $(s_0,s_1,\ldots,s_{\tau},r_{\tau+1},r_{\tau+2},\ldots,r_{n-1})$ such that
\[
r_j  = \oplus p_{j}|_{-1}(s)   
\]
for each $j ‌ \in \{n-1,n-2,\ldots,\tau+1\}$ and $p_j$ is defined by (\ref{pr}), then $N_F$ and $N_G$ generate the same output sequence.
\begin{flushright}
$\Box$
\end{flushright}

Since the functions $g_{n-1}, g_{n-2}, \ldots, g_{\tau}$ of a uniform Galois NLFSR 
depend on variables 
with indexes between 0 to $\tau$ only, the following property follows directly from the Theorem~\ref{init_state}.

\begin{lemma} \label{l1}
Let $N_F$ be an $n$-bit Fibonacci NLFSR 
and $N_G$ be an equivalent uniform Galois NLFSR with the terminal bit $\tau$.
If both $N_F$ 
and $N_G$ are initialized to any state $(s_0,s_1,\ldots,s_{n-1})$  such that $s_i = 0$ for all $i \in \{0,1,\ldots,\tau\}$, 
then they generate the same output sequence.
\end{lemma}

As an example, consider the 4-bit Fibonacci NLFSR $N_3$ with the feedback functions:
\[
\begin{array}{l}
f_3 = x_0 \oplus x_1 \oplus x_2 \oplus x_1 x_2 \\
f_2 = x_3  \\
f_1 = x_2 \\
f_0 = x_1 
\end{array}
\]
which is equivalent to the Galois NLFSRs $N_1$ and  $N_2$ from the previous example. 
The 3rd column of Table~\ref{ex} shows the sequence of states of $N_3$.
The terminal bits of $N_1$ and $N_2$ are 2 and 1, respectively. Therefore, is 
$(1000)$ is used as an initial state (2nd row of Table~\ref{ex}), all three NLFSRs 
generate the same output sequence $000101101001111$.



\section{Conclusion} \label{conc}

In this paper, we establish a relation between sequences of states 
generated by two equivalent NLFSRs and show how to compute the initial 
state for the Galois configuration 
which matches a given initial state of the Fibonacci configuration.

Many fundamental problems related to NLFSRs remain open. Probably the most important one
is finding a systematic
procedure for constructing NLFSRs with a guaranteed long period. 
Available algorithms either consider some special cases~\cite{JaS04}, or
applicable to small NLFSRs only~\cite{Fr82}.
The general 
problem is hard because there seems to be no simple algebraic theory
supporting it. Specifically, so far no analog of a primitive generator polynomial
has been found for the nonlinear case. 

\bibliographystyle{ieeetr}
\bibliography{bib}

\begin{thebibliography}{10}

\bibitem{Golomb_book}
S.~Golomb, {\em Shift Register Sequences}.
\newblock Aegean Park Press, 1982.

\bibitem{Ca05}
A.~Canteaut, ``Open problems related to algebraic attacks on stream ciphers,''
  in {\em WCC}, pp.~120--134, 2005.

\bibitem{robshaw94stream}
M.~Robshaw, ``Stream ciphers,'' Tech. Rep. TR - 701, July 1994.

\bibitem{AbBF94}
M.~Abramovici, M.~A. Breuer, and A.~D. Friedman, {\em Digital Systems Testing
  and Testable Design}.
\newblock Jon Willey and Sons, New Jersey, 1994.

\bibitem{Ma69}
J.~Massey, ``Shift-register synthesis and bch decoding,'' {\em IEEE
  Transactions on Information Theory}, vol.~15, pp.~122--127, 1969.

\bibitem{DuTT08}
E.~Dubrova, M.~Teslenko, and H.~Tenhunen, ``On analysis and synthesis of
  $(n,k)$-non-linear feedback shift registers,'' in {\em Design and Test in
  Europe}, pp.~133--137, 2008.

\bibitem{GaGK07}
B.~Gammel, R.~G{\"o}ttfert, and O.~Kniffler, ``Achterbahn-128/80: Design and
  analysis,'' in {\em SASC'2007: Workshop Record of The State of the Art of
  Stream Ciphers}, pp.~152--165, 2007.

\bibitem{hell-grain}
M.~Hell, T.~Johansson, and W.~Meier, ``{G}rain - a stream cipher for
  constrained environments,'' citeseer.ist.psu.edu/732342.html.

\bibitem{CHM05}
K.~Chen, M.~Henricken, W.~Millan, J.~Fuller, L.~Simpson, E.~Dawson, H.~Lee, and
  S.~Moon, ``Dragon: {A} fast word based stream cipher,'' in {\em eSTREM,
  {ECRYPT} Stream Cipher Project}, 2005.
\newblock Report 2005/006.

\bibitem{canniere-trivium}
C.~D. Canniere and B.~Preneel, ``{TRIVIUM} specifications,''
  citeseer.ist.psu.edu/734144.html.

\bibitem{cryptoeprint:2005:415}
B.~Gittins, H.~A. Landman, S.~O'Neil, and R.~Kelson, ``A presentation on {VEST}
  hardware performance, chip area measurements, power consumption estimates and
  benchmarking in relation to the aes, sha-256 and sha-512.'' Cryptology ePrint
  Archive, Report 2005/415, 2005.
\newblock http://eprint.iacr.org/.

\bibitem{GaGK06}
B.~M. Gammel, R.~G{\"o}ttfert, and O.~Kniffler, ``An {NLFSR}-based stream
  cipher,'' in {\em ISCAS}, 2006.

\bibitem{Du08}
E.~Dubrova, ``An equivalence preserving transformation from the {F}ibonacci to
  the {G}alois {NLFSR}s.''
\newblock http://arxiv.org/abs/0801.4079.

\bibitem{Sh09}
S.~Mansouri, {\em Re-Designing Grain Stream Cipher for Higher Throughput}.
\newblock M. Sc. Thesis, Royal Institutre of Technology (KTH), Sweden, 2009.

\bibitem{JaS04}
J.~S. I.~Janicka-Lipska, ``Boolean feedback functions for full-length nonlinear
  shift registers,'' {\em Telecommunications and Informatioin Technology},
  vol.~5, pp.~28--29, 2004.

\bibitem{Fr82}
H.~Fredricksen, ``A survey of full length nonlinear shift register cycle
  algorithms,'' {\em SIAM Review}, vol.~24, no.~2, pp.~195--221, 1982.

\end{thebibliography}

\end{document}